\documentclass[12pt]{iopart}
\newcommand{\boxed}[1]{#1}
\usepackage{graphicx}
\usepackage{color}
\begin{document}
\title{Adhesive contact of elastomers: effective adhesion energy and creep function}
%
\author{E. Barthel$^1$\footnote{etienne.barthel@saint-gobain.com} and C. Fr\'etigny$^2$}
\address{1) Surface du Verre et Interfaces, CNRS/Saint-Gobain, UMR 125, 93330, Aubervilliers Cedex France.
and
2) Laboratoire de Physico-Chimie des Polymères et des Milieux Dispersés,
UMR CNRS 7615 Ecole de Physique et Chimie Industrielles (ESPCI), 10 rue Vauquelin, 75213 Paris, France
}
%
%
\begin{abstract}
For the adhesive contact of elastomers, we propose expressions to quantify the impact of viscoelastic response on effective adhesion energy as a function of contact edge velocity. The expressions we propose are simple analytical functionals of the creep response and should be suitable for experimental data analysis in terms of measured rheologies. We also emphasize the role of the coupling between local stress field at the contact edge and the macroscopic remote loading (far field). We show that the contrast between growing and receding contacts originates from the impact of viscoelastic response on coupling, while the separation process at the contact edge is similarly affected by viscoelasticity in both cases.
\end{abstract}



\section{Introduction}\label{SecIntro}
Technological applications of soft polymers are multifarious. An especially noteworthy property is their high compliance resulting in remarkable contact behaviours. This is why they have been used for about a century for tyre manufacturing, for example. As a result the adhesive contact of elastomers has been thoroughly explored over the last decades. The problem neatly combines mechanical response of the materials and interfacial properties~\cite{JKR,Maugis78,Bouissou02,Ebenstein06,Wahl06,Charrault09}. Over the years a very thorough theoretical understanding has been reached, to such an extent that adhesive elastomeric contacts are routinely used for the assessment of surface properties: the contact will be established and the force necessary to pull out the sphere monitored as a function of the (decreasing) contact radius. By applying the JKR model,
an evaluation of the adhesion energy~\cite{Chaudhury91,Deruelle95}, characteristic of the interface is obtained. However, beyond the JKR model, the adhesive contact of elastomers hides some more complex features.

\subsection{Receding contacts and enhancement of adhesion}
While he JKR model for elastic bodies involves the \emph{reversible} adhesion energy (denoted $w$ here), it is usually found that for the \emph{rupture} of elastomeric adhesive contacts, the measured adhesion energies are considerably larger than $w$. This enhancement results from some viscoelastic dissipation which occurs in the polymeric material but originates from the tensile cohesive stresses involved in the reversible adhesive process~\cite{Gent72,Andrews73}. The question is the quantitative evaluation of this dissipative contribution given the viscoelastic response of the elastomer.

In this respect, the elastomeric adhesive contact problem, and the closely connected \emph{viscoelastic crack} problem, have received a lot of attention over the years~\cite{Wahl06,Schapery75a,Schapery75b,Schapery75c,Greenwood81,Schapery89,Hui03,Greenwood04,Greenwood06,Greenwood07,Morishita08}.
To describe the viscoelastic dissipation at the edge of the contact, the standard technique is to introduce a \emph{cohesive zone} (Fig.~\ref{Fig_Couplings}) to include the reversible adhesive stresses as boundary conditions to the viscoelasticity problem. The argument is well-known: for a linear viscoelastic system, an infinitely sharp tip would lead to infinite strain rate -- and no dissipation --  whatever the crack tip velocity~\cite{Hui98}. In contrast, introducing the \emph{range} of the cohesive interactions, there appears a zone with finite size (a-c
in Fig~\ref{Fig_Couplings}) over which the interaction stresses
do work. Then the viscoelastic material experiences a finite strain
rate resulting in velocity dependent dissipation through the viscoelastic response.

The loose statement that the viscoelastic crack response scales with the
dissipative component of the linear mechanical response is often
found \cite{Charrault09,Maugis85}. A more quantitative result can
be derived from the formulation by Schapery and Greenwood
\cite{Schapery75a,Schapery89,Greenwood04}.
However this formulation has yet only been applied to idealized
materials. Indeed in practice the numerical calculations are
complex~\cite{RahulKumar00,Greenwood04}, so that approximate schemes have been
developed~\cite{Schapery75b,Schapery75c,Greenwood81,Christensen81}. Another theory which takes into account the dissipative processes generated by a cohesive zone has been proposed~\cite{Gennes96,Saulnier04}. Heralding the fact that there
is no such direct relation with the dissipative function, it has
become known under the denomination of the "viscoelastic trumpet". In the present paper we propose a new approximate expression accounting for the adhesion energy enhancement by viscoelastic dissipation in a receding contact and demonstrate that the form is simple to use and more accurate than some standard approximations.

\subsection{Growing contacts and suppression of adhesion}
\label{Sec_Elastic_Adhesive_Contact}
\begin{figure}
\begin{center}
\includegraphics[width=4.5in]{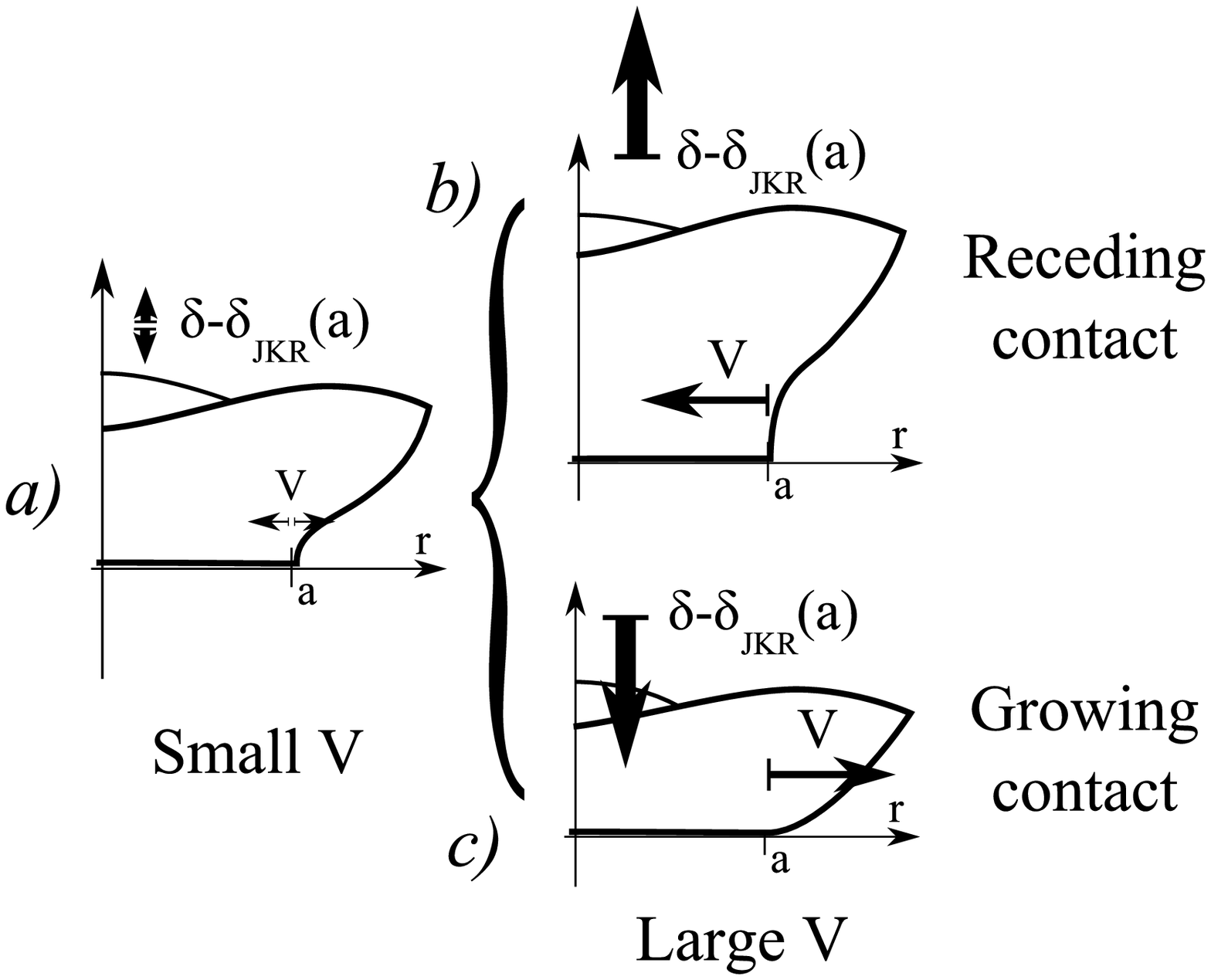}
\caption{Schematics of the adhesive contact for an {\it elastomer}.$\delta_{JKR}$ is the elastic JKR penetration for an adhesion energy $w$.
For small velocities, the system is close to equilibrium and receding and growing contacts are close to the elastic (JKR) limit with the relaxed modulus. For large velocities, strong deviations connected to the viscoelastic dissipative processes appear, which are quantified in the effective adhesion energy.}\label{Fig_Adhesive_Elastomer}
\end{center}
\end{figure}

An even more striking manifestation of viscoelastic dissipation in the adhesive contact of elastomers is the contrast between growing and receding contacts. As just mentioned above, for contact rupture (a receding contact), an increased adhesion energy is recorded. However, during loading, for an adhesive contact growing fast, the contact morphology is similar to an adhesionless (Hertz) contact (Fig.~\ref{Fig_Adhesive_Elastomer} c) and the equilibrium JKR-like contact (Fig.~\ref{Fig_Adhesive_Elastomer} a) is very slowly reached when the penetration, or the load, is held constant: for an elastomer which equilibrates instantaneously in a bulk mechanical test, it typically requires ten minutes for an adhesive contact to reach equilibrium~\cite{Vallet02}. This latter effect is less known because the suppression of adhesion is somehow less appreciated than its enhancement.

\begin{figure}
\begin{center}
\includegraphics[width=4.5in]{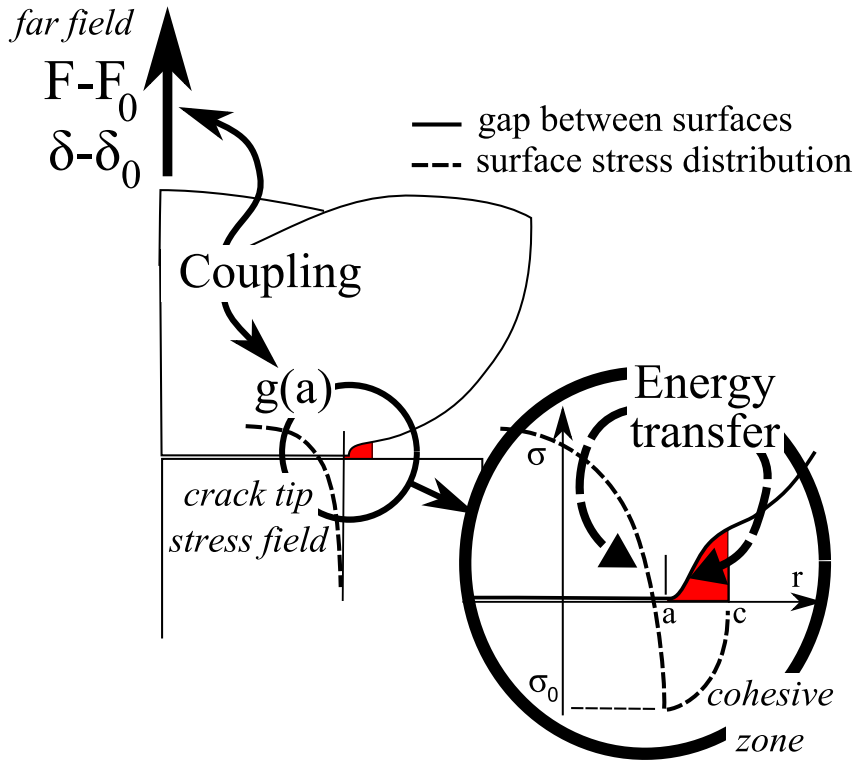}
\caption{Schematics of the macroscopic contact and the cohesive zone along with the coupling between the far field on the one hand, and the energy transfer within the cohesive zone on the other hand.}\label{Fig_Couplings}
\end{center}
\end{figure}

However, the loss of adhesion signals another interesting question. Indeed the cohesive zone process determines the local stress field at the the edge of the contact (or crack tip); the latter in turn must be matched with the far field, a process we call \emph{coupling} (Fig.~\ref{Fig_Couplings}). As we will show in this paper, the issue of the coupling in the adhesive contact of elastomers is simple for receding contacts, but not for growing contacts. For a growing contact, it was first qualitatively discussed by Johnson and Greenwood~\cite{Greenwood81}, then more explicitly dealt with by  Schapery~\cite{Schapery89} at various degrees of approximation. These results where subsequently adopted by Johnson and Greenwood~\cite{Johnson00,Greenwood04}.
In the present paper, we explicitly exhibit simple approximate expressions for the coupling between local contact edge stress field and remote, far field, loading. For a growing contact, the velocity dependence of the coupling demonstrates how the viscoelastic process quells the adhesion by smothering the coupling.

In brief, the emerging picture is that the measured adhesion energy is an
\emph{effective adhesion energy} (or apparent adhesion energy or apparent
work of adhesion) and that this effective adhesion energy (denoted $G_{eff}$ here)
may be suppressed (growing contact) or enhanced (receding contacts) as a result of viscoelastic dissipation.

\subsection{Strategy -- Effective adhesion energy, viscoelastic materials and elastomers}

For an elastomer, a soft polymeric material which does not flow (rubber-like material), the mechanical response is controlled by the molecular chain entropy~\cite{Treloar75}. At longer experimental time scales, elastomers exhibit a non-dissipative response. At smaller time scales, the viscoelastic behavior associated with the polymeric glass transition appears. However, as the glass transition temperature of elastomeric materials is low (typically below -30~$^\circ$C), and because of the time-temperature superposition principle, the viscoelastic time scale is far much shorter than the second at room temperature. As a result the rationale behind the use of the JKR model for \emph{elastomers} is as follows: the \emph{bulk} of the system effectively behaves as an elastic solid and assumes the response of the relaxed material, while viscoelastic effects act at the local scale, near the contact edge, inducing dissipation~\cite{Johnson00}.

This picture gives a a firm ground to the notion of effective adhesion energy. Let us consider two geometrically identical systems, one ideally elastic, and the other one elastomeric; and, in addition, that the long time modulus of the elastomeric material is equal to the modulus of the perfectly elastic material. Due to dissipation of the elastomer at the contact edge, identical far fields (same penetration, same applied force) require different reversible adhesion energies for the elastic system: a larger for the receding contact, a smaller for the growing contact. Conversely the effective adhesion of the elastomeric system is equal to the reversible adhesion energy which is necessary to produce identical far fields in the elastic system.

%
%


In the present paper we revisit the theory of the adhesive contact of elastomers from a different viewpoint.
Indeed there exists another set of relevant contact theories, which deal with the \emph{fully} viscoelastic systems, in which the contact timescale is of the order of the stress relaxation time, so that the process of stress relaxation in the bulk must be explicited. By \emph{restricting} such theories to elastomeric behaviour -- the elastic bulk approximation-- a consistent picture of viscoelastic adhesive contacts may be elaborated.

For the full viscoelastic contact problem, the solution proposed by Falsafi~\cite{Falsafi97} was later criticized by Hui who put forth a more consistent theory~\cite{Hui98} for both growing and receding contacts. Other solutions were later proposed based on the powerful Hankel transform technique~\cite{Yu90,Barthel99,Huguet00,Perriot04,Gacoin06,OnurSergici06} pioneered by Sneddon~\cite{Sneddon50}. Indeed, some years ago, using the Sneddon method, we have proposed another theory for the viscoelastic adhesive contact~\cite{Barthel02,Haiat03,Haiat07}. Two main results were obtained: 1) an analytical expression for the viscoelastic crack response; 2) a general relation between local crack tip stress field and far field relevant for the coupling problem in a fully viscoelastic system.

It is this theory which is the starting point for the present investigation of elastomeric adhesive contacts where we propose a new formulation for the effective adhesive response of elastomeric materials. For both growing and receding contacts, expressions are identified for the viscoelastic crack response (Eq.~\ref{Eq_Cohesive_Zone_Viscolastic}) and for the effective adhesion  (Eqs.~\ref{Eq_Effective_Adhesion_Receding} and \ref{Eq_Effective_Adhesion_Growing}). These expressions are cast in the form of simple analytical functionals of the creep function (Eqs.~\ref{Eq_Effective_Compliance_Opening},~\ref{Eq_Effective_Compliance_Closing} and \ref{Eq_Phi_0}). We expect they will be useful in the analysis of experimentally determined effective adhesion energies based on the knowledge of the creep function. The relations which these expressions bear to other elastomeric contact theories are also discussed.

\section{Model}
\subsection{Elastic Adhesive Contact}
We first briefly review the use of \emph{cohesive zones} in adhesive contact problems with \emph{purely elastic} systems.

\subsection{Cohesive zone -- Elastic material}\label{Sec_Cohesive_Zone}
In the cohesive zone the work of the tensile stresses during crack propagation induces a transfer between adhesion and mechanical energy~\cite{Barthel08}. It has been shown that there is little impact of the \emph{details} of the cohesive zone on the crack tip behaviour~\cite{Barthel98} so that the choice of a given cohesive zone model is not expected to significantly affect the results. The cohesive zone model may be chosen for convenience and while Greenwood~\cite{Greenwood04} used Maugis'~\cite{Maugis92} we prefer Greenwood's own "double Hertz" stress distribution~\cite{Greenwood98} (Fig.~\ref{Fig_Couplings}). Integrated over the cohesive zone size $\epsilon_0=c-a$, the work of the cohesive stresses ($\simeq \sigma_0$) in the surface displacements ($\simeq \sigma_0/{E}^\star$) can be calculated self-consistently. Here the elastic parameters are the Young's modulus $E$ and the Poisson ratio $\nu$, while ${{E}^\star}=E/(1-\nu^2)$ is the reduced elastic modulus. It is equal to the adhesion energy  $w$~\cite{Barthel98,Greenwood98}, so that for the double-Hertz model~\cite{Greenwood98}
\begin{equation}\label{Eq_Cohesive_Zone_Size_Elastic}
w=\frac{\pi}{4}\frac{{\sigma_0}^2\epsilon_0}{{E}^\star}\
\ \ \ \ \hbox{(elastic)}
\end{equation}
\subsection{Local stress field distribution}
The stress field at the edge of the contact zone (but inside,
$r<a$, Fig.~\ref{Fig_Couplings}) is controlled by the cohesive
zone stress field  ($a<r<c$). The resulting enhancement of the local tensile stress field is proportional to the stress intensity factor $K$. In the Sneddon formalism the form which appears naturally for the stress intensity factor is~\cite{Barthel02}:
\begin{equation}\label{Eq_Gdea}
g(a) \simeq -\frac{\pi}{4} \sigma_0 \sqrt{2a\epsilon_0}
\end{equation}
while $K=-2g(a)/\sqrt{\pi a}$.

The energy transfer between the physical process of adhesion and the local contact edge stress field (Eq.~\ref{Eq_Cohesive_Zone_Size_Elastic}) can be written
\begin{equation}\label{Eq_Cohesive_Zone_Elastic}
\boxed{w=\frac{2}{\pi}\frac{g(a)^2}{a{E}^\star}\ \ \ \ \ \hbox{(elastic)}}
\end{equation}
which we identify with the familiar expression for the energy release rate
\begin{equation}
w=\frac{K^2}{2{E}^\star}
\end{equation}

\subsection{Coupling with the far field}
In the elastic case, the coupling between the far field and the contact edge stress field is given by
\begin{equation}\label{Eq_Penetration_Elastic}
\boxed{\delta_{JKR} - \delta_H(a) = \frac{2}{E^\star} g(a)\ \ \ \ \ \hbox{(elastic)}}
\end{equation}
where $\delta_H(a)$ is the adhesionless contact radius. The coupling coefficient between contact edge stress field and far field is ${\cal C}= {2}/{E^\star}$. The force is
\begin{equation}\label{Eq_Force_Elastic}
F_{JKR}(a)=F_H(a)+4ag(a)={E^\star}\tilde F_H(a)+2a{E^\star}(\delta_{JKR}-\delta_H(a))
\end{equation}
where $F_H\equiv{E^\star}\tilde F_H$ is the adhesionless contact force. In anticipation of section~\ref{Sec_Viscoelastic_Adhesive_Contact}, $\tilde F_H$ is introduced because it contains the geometrical dependence of the (adhesionless) contact force and leaves out the mechanical response $E^\star$. Such a separation is possible for homogeneous half-spaces only~\cite{Perriot04}. Inserting Eq.~\ref{Eq_Cohesive_Zone_Elastic} the familiar JKR equations are recovered~\cite{JKR}.

\subsection{Far field -- flat punch displacement}
More precisely Eqs.~\ref{Eq_Penetration_Elastic} and
\ref{Eq_Force_Elastic} allow for an interpretation in terms of an
\emph{additional flat punch displacement} $\delta_{fp}(a)=\delta_{JKR}(a) - \delta_H(a)$
which produces the stress field singularity proportional to $g(a)$
at the crack tip~\cite{JKR,Barthel08}. Let us start from an
adhesionless contact ($g(a)=0$): the penetration $\delta$ is a function $\delta_H(a)$ of the contact radius $a$. Then we turn adhesion on and assume some adhesive process subsumed in $g(a)<0$. Eq.~\ref{Eq_Penetration_Elastic} expresses the fact that to keep
the contact radius constant while turning the adhesion on, one
needs to pull \emph{back} the sphere by a flat punch displacement
$\delta_{JKR}-\delta_H(a)$ proportional to $g(a)$. The resulting additional
force term in Eq.~\ref{Eq_Force_Elastic} is the flat punch force (the flat punch displacement times the flat punch stiffness
$2aE^\star$).

Alternatively, the same interpretation can be formulated as
follows: when adhesion is turned on for a given penetration
$\delta_{JKR}$, the contact radius increases so that
Eq.~\ref{Eq_Penetration_Elastic} is fulfilled. The cohesive
stresses tend to extend the contact zone in a manner similar to
wetting. This tendency for the cohesive stresses to expand the
contact zone clearly announces a possible role of \emph{creep} in
the coupling between local crack tip stress field and far field in
the case of viscoelastic materials (section~\ref{Sec_Coupling_Visco}).

\section{Viscoelastic Adhesive Contact}\label{Sec_Viscoelastic_Adhesive_Contact}
For a {\em linear viscoelastic} material~\cite{Lakes06}, following Schapery~\cite{Schapery75a} and Greenwood~\cite{Greenwood04} in the usual restrictions on the definition of the response for contact models, we introduce the creep function $\phi(t)$~\cite{Barthel02} which provides the delayed deformation in response to the history of applied stresses. Denoting  $E_\infty$ the \emph{relaxed modulus} (for the long time response) and ${E_0}$ the \emph{instantaneous modulus} (the short time response), we have ${\phi}(0)=2/{E_0}^\star$ and ${\phi}(+\infty)=2/{E_\infty}^\star$. $k$ is the contrast between relaxed and instantaneous moduli. An elastomeric material has $k\equiv {E_\infty}/{E_0}<<1$.


\subsection{Cohesive zone model, creep and effective crack tip compliance}
For a fast crack, the bonding or debonding processes will be effectively elastic and the modulus involved will be the instantaneous modulus. Similar considerations apply to the slow crack, where the relevant modulus will be the relaxed modulus. The latter is smaller by a factor $k$ resulting, for identical cohesive stresses, in larger deformations and proportionally smaller
cohesive zone size for a slow crack (Eq.~\ref{Eq_Cohesive_Zone_Elastic})~\footnote{Our intuition tends to correlate sharp necking on the crack edge with small cohesive zones. This is an error: the necking is a result of the far field while the cohesive zone size is relevant locally at the crack tip. Indeed the cohesive zone size $\epsilon$ is {\it directly} proportional to the square of the stress intensity factor $g(a)$ (Eq.~\ref{Eq_Gdea_Viscoelastic}). Note also that for an elastomeric contact, $g(a)$ only indirectly drives the flat punch displacement $\delta-\delta_H(a)$, as will be shown below (Eqs.~\ref{Eq_Penetration_Viscoelastic_Growing} and \ref{Eq_Penetration_Viscoelastic_Receding}). As a result for similar cohesive zone sizes, a strong (Fig.~\ref{Fig_Adhesive_Elastomer} b) or nearly absent (Fig.~\ref{Fig_Adhesive_Elastomer} c) necking may be recorded at the {\it macroscopic} scale.}.
More generally, for a viscoelastic system, the cohesive zone size $\epsilon$
depends upon crack tip velocity $V$ and for intermediate velocities,
the crack behavior will be characterized by some \emph{effective crack tip compliance} of the material (denoted $\phi_1$ here) and an intermediate value of the cohesive zone size. This is what a viscoelastic crack model must quantify.

Let us define the \emph{dwell time} $t_r$ (or transit time) as the time
it takes for the moving crack to cross its own imprint $\epsilon=c-a$
(Fig.~\ref{Fig_Couplings}). In the quasi static crack tip
approximation we assume the tip shape is convected, {\it i.e.} the
crack velocity $V$ does not change appreciably during a time
comparable with the dwell time $t_r$. We simply have $V=\epsilon
t_r$ and we want to calculate  $\epsilon(V)$, or rather $\epsilon(t_r)$.

In the viscoelastic case, we have shown~\cite{Barthel02} that the
crack tip processes can be described by the following effective crack tip compliance:
%
\begin{equation}\label{Eq_Cohesive_Zone_Size_Viscoelastic}
w=\frac{\pi}{8}{\sigma_0}^2\epsilon(t_r){\phi}_{1}(t_r)\ \ \ \ \ \hbox{(viscoelastic)}
\end{equation}
where ${\phi}_{1}(t_r)$ is either ${\phi}_{1,op}(t_r)$ or ${\phi}_{1,cl}(t_r)$ with
\begin{equation}\label{Eq_Effective_Compliance_Opening}
{\phi}_{1,op}(t)=\frac{2}{t^2}\int_0^t(t-\tau)\phi(\tau)d\tau\ \ \ \ \ \hbox{(opening crack)}
\end{equation}
and
\begin{equation}\label{Eq_Effective_Compliance_Closing}
\phi_{1,cl}(t)=\frac{2}{t^2}\int_0^t\tau {\phi}(\tau)d\tau\ \ \ \ \ \hbox{(closing crack)}\end{equation}

Locally the contact edge stress field, as characterized by the stress intensity factor $g(a)$
is~\cite{Barthel02}
\begin{equation}\label{Eq_Gdea_Viscoelastic}
g(a(t),t_r) \simeq -\frac{\pi}{4} \sigma_0 \sqrt{2a(t)\epsilon(t_r)}\end{equation}
%
As a result we have
\begin{equation}\label{Eq_Cohesive_Zone_Viscolastic}
\boxed{w=\frac{g(a)^2}{\pi a}\phi_1(t_r)\
\ \ \ \ \hbox{(viscoelastic)}}
\end{equation}
where $\phi_1(t_r)={\phi}_{1,op}(t)$ or ${\phi}_{1,cl}(t)$ for an opening or closing crack.

For practical purposes Eqs.~\ref{Eq_Effective_Compliance_Opening} and~\ref{Eq_Effective_Compliance_Closing} for the effective crack tip compliance $\phi_1(t_r)$ may well be the main result of the paper. However, to be of use,
it must first be connected to the measurable parameter, the effective adhesion, a task we complete only in section~\ref{Sec_Effective_Adhesion}.

Here we will discuss some representative quantitative results for
the effective crack tip compliances
Eqs.~\ref{Eq_Effective_Compliance_Opening}
and~\ref{Eq_Effective_Compliance_Closing}. We can visualize the effective crack tip compliance by  its impact on the cohesive
zone size, which is directly (Eqs.~\ref{Eq_Cohesive_Zone_Size_Elastic} and~\ref{Eq_Cohesive_Zone_Size_Viscoelastic})
%
\begin{equation}\label{Eq_Cohesive_Zone_Size_Normalized_Viscoelastic}
\epsilon(t_r) = \frac{2}{{E_0}^\star}\frac{\epsilon_0}{{\phi}_{1}(t_r)}
\end{equation}

\begin{figure}
\begin{center}
\includegraphics[width=3.25in]{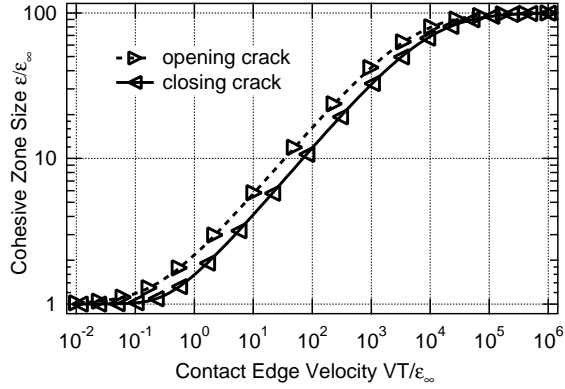}
\caption{Cohesive zone size as a function of velocity for opening and closing cracks. Creep follows an exponential time dependence (three-element model) with $k=0.01$. The viscoelastic cohesive zone processes are nearly identical for closing or opening cracks. }\label{Fig_Cohesive_Zone_Size}
\end{center}
\end{figure}

We consider the special case where the creep function is a three-element model (\ref{Sec_Appendix_Creep_Function}), which is frequently used as a coarse rendering of viscoelastic behavior. $T$ is the creep time and the ratio of long time to instantaneous elastic modulus is $k=0.01$. The normalized cohesive zone sizes for opening or closing cracks are plotted on
Fig.~\ref{Fig_Cohesive_Zone_Size}. The results are expressed as a function of the "fast crack" characteristic velocity $V=\epsilon_0/T$. As expected Eqs.~\ref{Eq_Effective_Compliance_Opening} and
\ref{Eq_Effective_Compliance_Closing} predict a transition between the relaxed cohesive zone at low velocity and the instantaneous cohesive zone at high velocity. The transition occurs around a normalized velocity of 1. The minor offset between the opening and the closing cracks is simply due to the different cohesive stress history experienced by the cracks moving in opposite directions. Altogether however the two cases are very similar and the curves collapse when a scaling by a factor of about 2 is applied on the velocity.

\begin{figure}
\begin{center}
\includegraphics[width=3.25in]{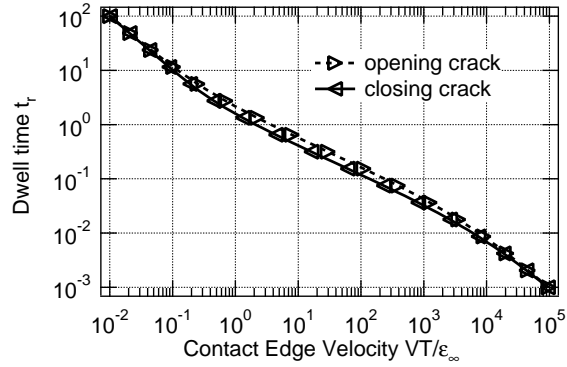}
\caption{Dwell time as a function of velocity for opening and closing cracks. Creep follows an exponential time dependence (three-element model) with $k=0.01$. }\label{Fig_Dwell_Time}
\end{center}
\end{figure}


\subsection{Coupling between local and far field}\label{Sec_Coupling_Visco}
\subsubsection{Penetration}
For a growing contact, we found that (Eq. 14 in \cite{Barthel02})
%
\begin{equation}\label{Eq_Penetration_Viscoelastic_Growing}
\boxed{\delta(t)-\delta_H(a(t)) \simeq \phi_0(t_r) g(a(t),t_r)\ \ \ \ \ \hbox{(growing contact)}}
\end{equation}

where
%
\begin{equation}\label{Eq_Phi_0}
\phi_0(t)=\frac{1}{t}\int_0^td\tau \phi(t-\tau)\end{equation}

Of course, since the contact is growing, it is the expression for a closing crack (Eq.~\ref{Eq_Effective_Compliance_Closing}) which must be taken for $\phi_1$ in Eq.~\ref{Eq_Cohesive_Zone_Viscolastic}.

At this level of approximation, for a growing crack, the coupling coefficient between
crack tip stress field and far field is
${\cal C}=\phi_0(t_r)$. The coupling coefficient directly reflects the creep acting over the cohesive zone during time $t_r$ and its impact on the penetration. As the effective crack tip compliance, the coupling coefficient exhibits a strong variation (Fig.~\ref{Fig_Coupling_Coefficient}) with dwell time $t_r$ (or conversely with crack velocity, Fig.~\ref{Fig_Dwell_Time}).
If $t_r$ is small - or the contact edge velocity large -, creep is inefficient and
${\cal C}\simeq 2/{E^\star}_0$: the additional flat punch displacement is smaller by a factor $k$ than the flat punch displacement for a relaxed material (Fig.~\ref{Fig_Adhesive_Elastomer} c)~\footnote{It is the flat punch \emph{displacement} for an elastic system with the \emph{instantaneous} modulus. Of course, the force (Eq.~\ref{Eq_Force_Viscoelastic_Growing}) is \emph{not} the force for the instantaneous modulus.}: the necking characteristic of the JKR adhesive contact is nearly absent and contact growth appears to proceed as a virtually non-adhesive contact. If $t_r$ is large - or contact edge velocity small -, ${\cal C}\simeq 2/{E^\star}_\infty$ and the additional flat punch displacement is large: we revert to the completely elastic, relaxed, adhesive contact (Fig.~\ref{Fig_Adhesive_Elastomer} a).

For a receding contact
\begin{equation}\label{Eq_Penetration_Viscoelastic_Receding}
\boxed{\delta(t) - \delta_H(a(t)) \simeq \frac{2}{{E^\star}_{\infty}} g(a(t),t_r)\ \ \ \ \ \hbox{(receding contact)}}
\end{equation}
where $\phi_1$ is for an opening crack (Eq.~\ref{Eq_Effective_Compliance_Opening}) since the contact is receding.
Eq.~\ref{Eq_Penetration_Viscoelastic_Receding} turns out to be the equation for the elastic case
(Eq.~\ref{Eq_Penetration_Elastic}) with the relaxed modulus. The
coupling coefficient between crack tip stress field and far field
is ${\cal C}={2}/{{E^\star}_{\infty}}$, which is independent from $t_r$ (Fig.~\ref{Fig_Coupling_Coefficient}).

For a receding crack, creep plays no role in the coupling. The surface displacements resulting from creep occur outside of the contact zone and move away from it so that they will not affect the far field. The coupling between far field and contact edge stress field is as in the usual JKR model (Fig.~\ref{Fig_Coupling_Coefficient}). Of course creep still plays a role in the (independent) self-consistent crack tip problem (section~\ref{Sec_Cohesive_Zone}), which determines the effective crack tip compliance.

\subsubsection{Force}
To complete the picture we also study the force. Expressions for the force are more complex because they result from the spatial integral of the normal surface stress. In both cases, since the bulk is fully relaxed, the first term on the RHS is the Hertz force with the relaxed modulus. The additional flat punch term is the product of the (relaxed~) contact stiffness with the actual flat punch displacements $\delta(t) - \delta_H(a(t))$ as given by~Eqs.~\ref{Eq_Penetration_Viscoelastic_Growing} and \ref{Eq_Penetration_Viscoelastic_Receding} respectively.

For a growing contact:
\begin{equation}\label{Eq_Force_Viscoelastic_Growing}
F(t) \simeq {E^\star_{\infty}}{\tilde
F}_H(a(t))+ \frac{{E^\star}_{\infty}}{2}\phi_0(t_r) 4 a g(a(t))
\end{equation}
Note that the flat punch force term is controlled by the specific coupling coefficient for the growing contact.

For a receding contact:
\begin{equation}\label{Eq_Force_Viscoelastic_Receding}
F(t) \simeq
{{E^\star}_{\infty}}{\tilde F}_H(a(t)+ 4 a(t)
g(a(t))\end{equation} which is the force equation for the elastic
case, with the relaxed modulus.

\subsection{Effective adhesion}\label{Sec_Effective_Adhesion}
We now consider the \emph{effective adhesion} $G_{eff}$ as defined in section~\ref{Sec_Elastic_Adhesive_Contact}. It can be calculated comparing Eqs.~\ref{Eq_Penetration_Viscoelastic_Growing} and \ref{Eq_Penetration_Viscoelastic_Receding} with Eq.~\ref{Eq_Penetration_Elastic}, or Eqs.~\ref{Eq_Force_Viscoelastic_Growing} and \ref{Eq_Force_Viscoelastic_Receding} with \ref{Eq_Force_Elastic}. In both cases $g(a)$ is determined by Eqs.~\ref{Eq_Cohesive_Zone_Viscolastic} or \ref{Eq_Cohesive_Zone_Elastic} as appropriate.

For the receding contact (using the opening crack expression for the effective crack tip compliance Eq.~\ref{Eq_Effective_Compliance_Opening}) the
effective adhesion is simply the normalized cohesive size
\begin{equation}
\frac{{\cal G}_{eff}}{w} = \frac{\epsilon}{\epsilon_\infty}
\end{equation}
so that
%
\begin{equation}\label{Eq_Effective_Adhesion_Receding}
\boxed{\frac{{\cal G}_{eff}}{w} = \frac{\phi_{1,op}(\infty)}{\phi_{1,op}(t_r)}\ \ \ \ \ \ \ \ \hbox{(receding contact)}}
\end{equation}
where $\phi_{1,op}(\infty)=2/{E^\star}_{\infty}$.
For contact growth (with the closing crack expression for the effective crack tip compliance Eq.~\ref{Eq_Effective_Compliance_Closing}), due to the non trivial coupling coefficient
${\cal C}=\phi_0(t_r)$
%
\begin{equation}
\frac{{\cal G}_{eff}}{w}=\left(\frac{{E^\star}_{\infty}}{2}\right)^2\phi_0^2(t_r)\frac{\epsilon}{\epsilon_\infty}
\end{equation}

so that
%
\begin{equation}\label{Eq_Effective_Adhesion_Growing}
\boxed{\frac{{\cal G}_{eff}}{w}=\frac{\phi_0^2(t_r)}{\phi_{1,cl}(\infty)\phi_{1,cl}(t_r)}\ \ \ \ \ \ \ \ \hbox{(growing contact)}}
\end{equation}
where $\phi_{1,cl}(\infty)=2/{E^\star}_{\infty}$.

\begin{figure}
\begin{center}
\includegraphics[width=3.25in]{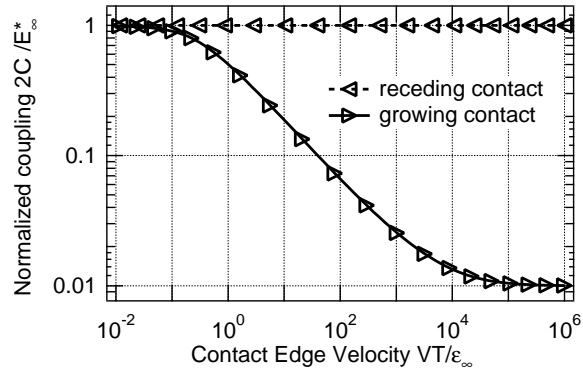}
\caption{Normalized coupling coefficient {\cal C} between crack tip stress
field and far field as a function of contact zone velocity for
growing and for receding
contacts. Creep follows an exponential time dependence (three-element model) with $=k=0.01$.}\label{Fig_Coupling_Coefficient}
\end{center}
\end{figure}

\begin{figure}
\begin{center}
\includegraphics[width=3.25in]{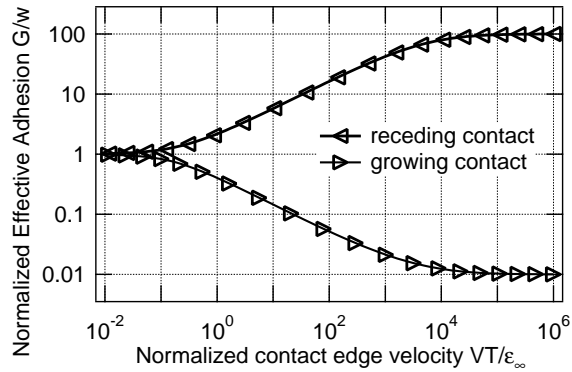}
\caption{Effective adhesion energy for growing and for receding
contacts. The nearly identical crack tip processes
(Fig.~\ref{Fig_Cohesive_Zone_Size}) result in very different effective
adhesion energies because of the dependence of the far field/local
field coupling coefficient with contact edge velocity
(Fig.~\ref{Fig_Coupling_Coefficient}). The result of the product
rule is also shown (see~\ref{Sec_Appendix_Effective_Adhesion}).}\label{Fig_Effective_Adhesion}
\end{center}
\end{figure}

Typical results are plotted on Fig.~\ref{Fig_Effective_Adhesion} for a three-element model with $k=0.01$. Despite nearly identical cohesive zone responses for a given velocity (Fig.~\ref{Fig_Cohesive_Zone_Size}), the effective adhesion energies for growing and receding contacts separate at higher velocities to reach their respective limit values equal to $k$ and $1/k$ respectively. This contrast is due to the strong decrease of the coupling coefficient for growing contacts at higher velocities (Fig.~\ref{Fig_Coupling_Coefficient}).

\section{Discussion}
With Eqs.~\ref{Eq_Effective_Adhesion_Receding} and~\ref{Eq_Effective_Adhesion_Growing} we have provided expressions for the effective adhesion of elastomeric contacts for receding and for growing contacts. These expressions are very general and can easily be applied to arbitrary creep functions. We now compare our results to predictions and approximations of the literature.

\begin{figure}
\begin{center}
\includegraphics[width=3.25in]{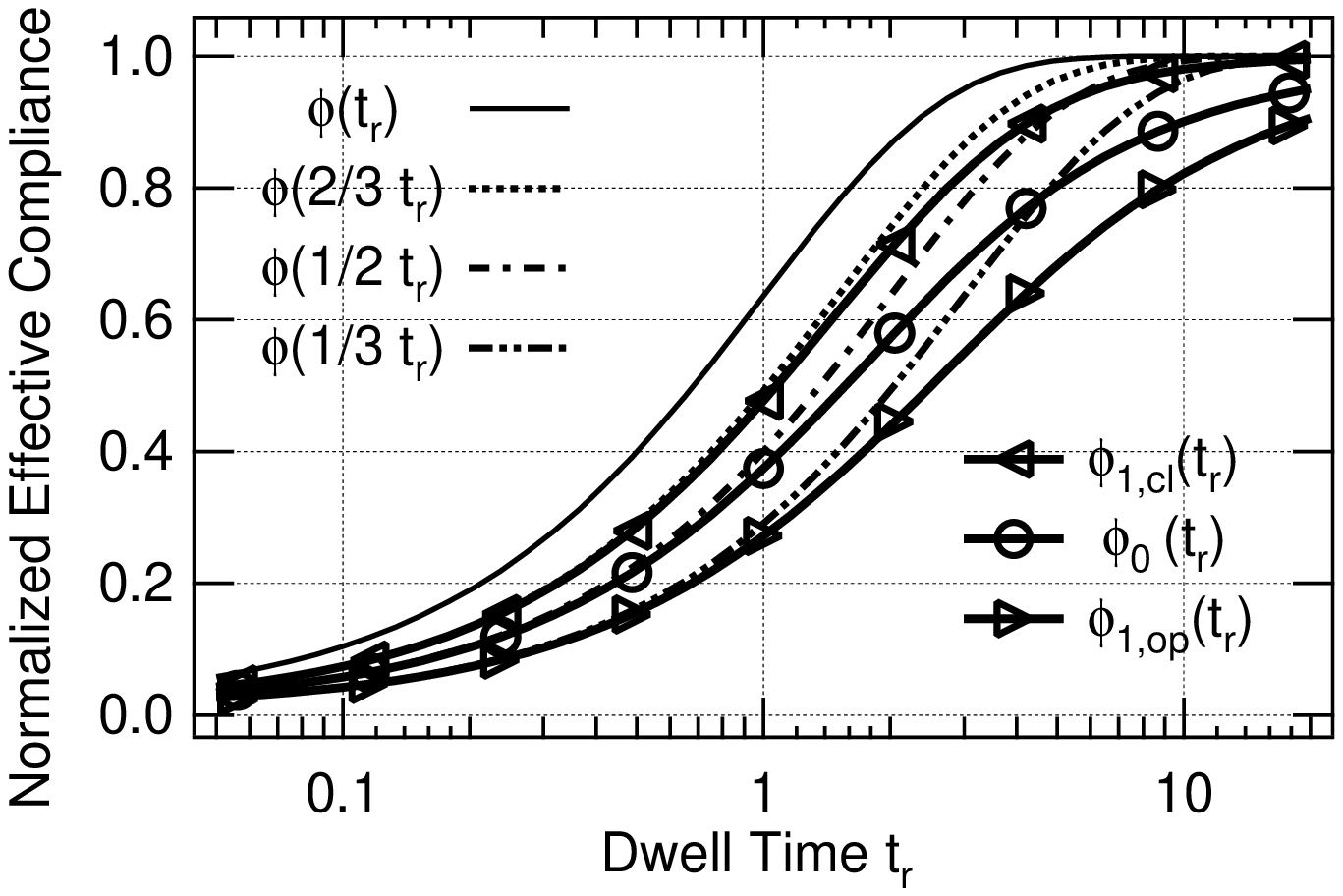}
\caption{Normalized effective compliances $\phi_1$ for opening and closing cracks as predicted by Eqs.~\ref{Eq_Effective_Compliance_Opening} and \ref{Eq_Effective_Compliance_Closing}), and comparison with the creep function approximation~\cite{Greenwood81} ($k=0.01$). Our results match the creep function approximation at shorter dwell times (high contact edge velocities), where it is known to be valid, and also captures the deviations evidenced at longer dwell times (slower velocitites) known from the earlier numerical calculations~\cite{Greenwood04}. The coupling constant $\phi_0$ as predicted by Eq.~\ref{Eq_Phi_0} and its short time expansion are also displayed. }\label{Fig_Effective_Compliance}
\end{center}
\end{figure}
%
%
\subsection{Receding contact and effective adhesion energy}
%
%
Based on earlier work by Schapery~\cite{Schapery75b}, Greenwood and Johnson~\cite{Greenwood81} have proposed a simple approximation. In this \emph{creep function approximation} the effective crack tip compliance is equal to the creep function evaluated at some \emph{effective dwell time} $t^*$. By assumption the effective dwell time $t^*$ is {\it proportional} to the dwell time $t_r$, that is $t^\star=p t_r$ for an opening crack.
The approximation is completely defined if the proportionality coefficient $p$
is specified. A value of $1/3$ has been proposed~\cite{Schapery75a,Greenwood81}. In 2004 Greenwood performed numerical calculations to improve the evaluation of the effective dwell time $t^\star$. He proposed $p\simeq1/4$ for a three-element viscoelastic solid with $k=0.02$ and a Dugdale-Maugis cohesive zone~\cite{Maugis92}.

In the same paper~\cite{Greenwood04} he also showed that the creep function approximation is only valid at shorter dwell times (or high contact edge velocity): at longer dwell times (or low contact edge velocity), the creep function approximation for the effective crack tip compliance is inaccurate and an effective dwell time can no longer be defined.

Here we have proposed a different type of expression for the effective crack tip compliances (Eqs.~\ref{Eq_Effective_Compliance_Opening} and~\ref{Eq_Effective_Compliance_Closing}). Compared to Greenwood's creep function approximation, which is based on the creep function itself, but evaluated at an effective dwell time, our expressions emerge from higher order developments and assume the form of moments of the creep function. For the three-element model, our effective crack tip compliances can be explicited analytically (Sec.~\ref{Sec_Appendix_Creep_Function}). To compare with the creep function approximation, $\phi_{1,op}$
is plotted on Fig.~\ref{Fig_Effective_Compliance}. We find that this function tracks the creep function at short dwell times (or high contact edge velocities), allowing for explicit evaluation of the effective dwell time. Expansion of 
Eq.~\ref{Eq_Appendix_Effective_Compliance_1cl} shows that $p=1/3$.
This result is (accidentally) identical to Schapery's~\cite{Schapery75b}, and similar to Greenwood's~\cite{Greenwood04}. Indeed explicit values may differ due to different cohesive zone types~\cite{Greenwood07} and different models for the viscoelastic response.

However what also clearly appears (Fig.~\ref{Fig_Effective_Compliance}) is that for longer dwell times (or lower contact edge velocities) the effective crack tip compliances converge more slowly to their asymptote than the creep function. This result is in full agreement with the trend identified by Greenwood using extensive numerical calculations~\cite{Greenwood04} and demonstrates the improved validity of the higher order expansion. We believe that our approximate expressions retain most of the simplicity of the creep function approximation but provide better evaluations of the effective response especially at low contact edge velocities.

\subsection{Growing contact and effective adhesion energy}\label{Sec_Power_Law}
It is apparent that the viscoelastic processes at the contact edge are very similar for receding and for growing cracks (Fig.~\ref{Fig_Cohesive_Zone_Size}. Indeed, in terms of effective adhesion energy, our results (Eqs.~\ref{Eq_Penetration_Viscoelastic_Growing} and \ref{Eq_Penetration_Viscoelastic_Receding}) demonstrate that the difference between receding and growing contacts (Fig.~\ref{Fig_Effective_Adhesion}) lies in the coupling (Fig.~\ref{Fig_Coupling_Coefficient}), not directly the contact edge viscoelastic crack response (\ref{Sec_Appendix_Effective_Adhesion}). At higher velocities, nearly identical viscoelastic crack responses for growing or receding contacts result in very contrasted effective adhesions because the coupling coefficients between local stress field and far field (Fig.~\ref{Fig_Coupling_Coefficient}) differ by a factor $k$ depending on the direction of contact edge motion. Similar results may be found in the literature, especially in~\cite{Schapery89} (Eqs.~63) and~\cite{Greenwood04} (par. 3 and appendix) but we believe that stronger emphasis on the coupling, such as conveyed by Eq.~\ref{Eq_Effective_Adhesion_Growing} and Fig.~\ref{Fig_Coupling_Coefficient}, best  brings out the workings of the growing contact.
%
%
%
%
%
\section{Conclusion}
We have proposed approximate analytical expressions (Eqs.~\ref{Eq_Effective_Adhesion_Receding} and \ref{Eq_Effective_Adhesion_Growing}) for the effective adhesion energy of elastomers. These expressions are easy to use because they involve simple functionals (Eqs.~\ref{Eq_Effective_Compliance_Opening}, \ref{Eq_Effective_Compliance_Closing} and \ref{Eq_Phi_0}) of the creep function.

We have shown that our expressions match the existing approximate model, the creep function approximation, for high contact edge velocities, and improve on it at low contact edge velocities. In this regime, our model correctly predicts the deviation of the compliance from the creep function approximation which has been observed numerically.

We also propose a consistent form for the effective adhesion energy of growing contacts, where emphasis is laid on the role of the coupling between local stress field and far field. In particular it very clearly brings out the similarity between the cohesive zone processes for opening or closing cracks while the contrasted effective adhesion energies are fully ascribed to the asymmetry found in the coupling.

We believe these results pave the way to more extensive experimental comparisons between bulk viscoelastic response and effective adhesion of elastomers.

\appendix
\section{Three-element model -- Creep function}\label{Sec_Appendix_Creep_Function}
The three-element model is an exponential creep function
\begin{equation}
\phi(t) = \frac{2}{{E_\infty}^\star}\left(1-(1-k)\exp{(-t/T)}\right)
\end{equation}

For the three-element model,
$\phi_0(t)$, $\phi_{1,op}(t)$ and $\phi_{1,cl}(t)$ can be calculated analytically from Eqs.~\ref{Eq_Phi_0}, \ref{Eq_Effective_Compliance_Opening} and \ref{Eq_Effective_Compliance_Closing}, giving
\begin{equation}\label{Eq_Appendix_Effective_Compliance_0}
\phi_0(t) = \frac{2}{{E_\infty}^\star}\left(1-\frac{1-k}{t/T}\left\{1-\exp{(-t/T)}\right\}\right)
\end{equation}
\begin{equation}\label{Eq_Appendix_Effective_Compliance_1op}
\phi_{1,cl}(t) = \frac{2}{{E_\infty}^\star}\left(1-2\frac{1-k}{(t/T)^2}\left\{1-(1+t/T)\exp{(-t/T)}\right\}\right)
\end{equation}
\begin{equation}\label{Eq_Appendix_Effective_Compliance_1cl}
\phi_{1,op}(t) = \frac{2}{{E_\infty}^\star}\left(1+2\frac{1-k}{(t/T)^2}\left\{1-t/T-\exp{(-t/T)}\right\}\right)
\end{equation}

\section{Effective adhesion and effective dwell time}\label{Sec_Appendix_Effective_Adhesion}
It has been proposed~\cite{Greenwood81,Schapery89,Johnson00} that the effective adhesion can be expressed as

\begin{equation}\label{Eq_G_eff_rec_GJ}
    \frac{G_{eff}}{w}=\frac{2}{{E_\infty}^\star\phi(t^\star)}\ \ \ \ \ \ \hbox{(receding contact)}
\end{equation}
and
\begin{equation}\label{Eq_G_eff_grow_GJ}
    \frac{G_{eff}}{w}=\frac{{E_\infty}^\star\phi({t_b}^\star)}{2}\ \ \ \ \ \ \hbox{(growing contact)}
\end{equation}
where $\phi(t^\star)$ and $\phi({t_b}^\star)$ are effective compliances for the crack tip creep process during contact edge propagation, for opening and closing cracks.

A linear theory of the coupling between local crack tip stress field and far field should involve the crack tip process in a similar way, irrespective of the contact edge direction (receding or growing). The fact that the effective crack tip compliance appears either at the numerator (Eq.~\ref{Eq_G_eff_rec_GJ}) or the denominator (Eq.~\ref{Eq_G_eff_grow_GJ}) when the crack direction is reversed is therefore puzzling.


Our equations Eqs.~\ref{Eq_Effective_Adhesion_Receding} and \ref{Eq_Effective_Adhesion_Growing} display a somewhat simpler structure. They explicitly call in the coupling $\phi_0(t_r)$ which results from the matching of the local stress field and the far field. Compared to Eq.~\ref{Eq_G_eff_grow_GJ}, this form restores the effective crack tip compliance at the denominator of
Eq.~\ref{Eq_Effective_Adhesion_Growing} so that the symmetry is preserved.
In this respect, note that the Schapery Equations~\ref{Eq_G_eff_grow_GJ} and \ref{Eq_G_eff_rec_GJ} were first derived under the assumption of a power law creep function and presumably better apply to that case, where these equations clearly become equivalent to Eqs.~\ref{Eq_Effective_Adhesion_Receding} and \ref{Eq_Effective_Adhesion_Growing}.

%
\newpage
\bibliographystyle{unsrt}
\bibliography{D:/data/Biblio_Listes/Mecanique,D:/data/Biblio_Listes/mecacouche,D:/data/Biblio_Listes/Fracture,D:/data/Biblio_Listes/Indentation,D:/data/Biblio_Listes/ContactAdhesion,D:/data/Biblio_Listes/ForcesdeSurface,D:/data/Biblio_Listes/MecaSolGel,D:/data/Biblio_Listes/Materiaux,D:/data/Biblio_Listes/Capillarity}
\end{document}